\documentclass{amsart}

\usepackage{amsmath,amscd,amssymb,amsthm}
\usepackage{enumerate,mathrsfs}
\usepackage{geometry}
\usepackage{hyperref}
\usepackage{url}
\usepackage{ifthen}
\usepackage[all]{xy} \xyoption{2cell} \UseAllTwocells

\numberwithin{equation}{section}
\newtheorem{theorem}{Theorem}[section]

\theoremstyle{plain}
\newtheorem{thm_}[theorem]{Theorem}
\newtheorem{lemma_}[theorem]{Lemma}
\newtheorem{prop_}[theorem]{Proposition}
\newtheorem{cor_}[theorem]{Corollary}
\newtheorem{eg_}[theorem]{Example}
\newtheorem{con_}[theorem]{Conjecture}
\newtheorem*{cons_}{Conjecture}

\theoremstyle{definition}
\newtheorem{thmu_}[theorem]{Theorem}
\newtheorem*{thmus_}{theorem}
\newtheorem{propu_}[theorem]{Proposition}
\newtheorem*{propus_}{Proposition}
\newtheorem{coru_}[theorem]{Corollary}
\newtheorem*{corus_}{Corollary}
\newtheorem{lemu_}[theorem]{Lemma}
\newtheorem*{lemus_}{Lemma}
\newtheorem{egu_}[theorem]{Example}
\newtheorem*{egus_}{Example}
\newtheorem{def_}[theorem]{Definition}
\newtheorem*{defs_}{Definition}
\newtheorem{rk_}[theorem]{Remark}
\newtheorem*{rks_}{Remark}
\newtheorem{ex_}[theorem]{Remark}

\newcommand{\thm}[1]{\begin{thm_}#1\end{thm_}}
\newcommand{\thmu}[1]{\begin{thmu_}#1\end{thmu_}}

\newcommand{\lemm}[1]{\begin{lemma_}#1\end{lemma_}}
\newcommand{\lemu}[1]{\begin{lemu_}#1\end{lemu_}}

\newcommand{\egu}[1]{\begin{egu_}#1\end{egu_}}

\newcommand{\prop}[1]{\begin{prop_}#1\end{prop_}}
\newcommand{\propu}[1]{\begin{propu_}#1\end{propu_}}

\newcommand{\defi}[1]{\begin{def_}#1\end{def_}}

\newcommand{\rk}[1]{\begin{rk_}#1\end{rk_}}

\newcommand{\cor}[1]{\begin{cor_}#1\end{cor_}}

\newcommand{\con}[1]{\begin{con_}#1\end{con_}}

\newcommand{\pf}[1]{\begin{proof}#1\end{proof}}

\DeclareMathOperator{\ord}{ord}

\DeclareMathOperator{\Gal}{Gal}

\newcommand{\QQ}{\mathbb Q}
\newcommand{\RR}{\mathbb R}
\newcommand{\ZZ}{\mathbb Z}
\newcommand{\fa}{\mathfrak a}%

\newcommand{\fo}{\mathfrak o}%
\newcommand{\p}{\mathfrak p}
\newcommand{\q}{\mathfrak q}
\newcommand{\A}{\mathfrak A}
\newcommand{\B}{\mathfrak B}

\newcommand{\fP}{\mathfrak P}%
\newcommand{\Q}{\mathfrak Q}

\newcommand{\s}{\sigma}

\newcommand{\tm}{\times}%

\newcommand{\ra}{\rightarrow}

\newcommand{\is}[2]{\xymatrix@-4mm{#1 \ar[r]^-{\sim} & #2 }}
\newcommand{\mis}[2]{\xymatrix@-2mm{#1 \ar[r]^-{\sim} & #2 }}

\newcommand{\dra}[4]{\xymatrix@-4mm{#1 \ar@<.5ex>[r]^-{#3} \ar@<-.5ex>[r]_-{#4}& #2 }}
\newcommand{\era}[5]{\xymatrix@-4mm{#1 \ar[r] &#2 \ar@<.5ex>[r]^-{#4} \ar@<-.5ex>[r]_-{#5}& #3 }}

\newcommand{\tu}[1]{\text{\upshape #1}}

\DeclareFontFamily{U}{wncy}{}
\DeclareFontShape{U}{wncy}{m}{n}{%
   <5>wncyr5%
   <6>wncyr6%
   <7>wncyr7%
   <8>wncyr8%
   <9>wncyr9%
   <10>wncyr10%
   <11>wncyr10%
   <12>wncyr6%
   <14>wncyr7%
   <17>wncyr8%
   <20>wncyr10%
   <25>wncyr10}{}
\DeclareMathAlphabet{\cyrille}{U}{wncy}{m}{n}

\newcommand{\eq}[1]{\begin{equation}#1\end{equation}}
\newcommand{\eqn}[1]{\begin{equation*}#1\end{equation*}}

\newcommand{\aln}[1]{\begin{align*}#1\end{align*}}

\newcommand{\enmt}[1]{\begin{enumerate}#1\end{enumerate}}

\newcommand{\tabl}[3]{\begin{center}\small{#1\\}\begin{tabular}{#2}#3\end{tabular}\end{center}}

\newcommand{\aci}[1]{\ar@{^(->}[#1]|-{/}}
\newcommand{\coaci}[1]{\ar@{_(->}[#1]|-{/}}
\newcommand{\aoi}[1]{\ar@{^(->}[#1]|-{\circ}}
\newcommand{\coaoi}[1]{\ar@{_(->}[#1]|-{\circ}}
\makeatletter
\def\citet@url@sp{https://stacks.math.columbia.edu/}
\def\citet@bib@sp{stacks-project}
\def\citet@url@kd{https://kerodon.net/}
\def\citet@bib@kd{kerodon}
\newcommand{\citet@tag}[2]{\href{#2tag/#1}{#1}}
\newcommand{\citet@taglist}[2]{%
 \def\@citet@e{}%
 \def\@citet@tag@n{0}
 \@for\@citet@tag:=#1\do{%
  \edef\@citet@tag@n{\the\numexpr\@citet@tag@n + 1}%
 }%
 \def\@citet@tags{%
  \def\@citet@tag@i{0}%
  \@for\@citet@tag:=#1\do{%
   \edef\@citet@tag@i{\the\numexpr\@citet@tag@i + 1}%
   \ifthenelse{\@citet@tag@i > 1}{
    \ifthenelse{\@citet@tag@i = \@citet@tag@n}{
     \citet@seplast%
    }{%
     \citet@sep%
    }%
   }{}%
   \citet@entry{\@citet@tag}{#2}
  }%
 }%
 \ifthenelse{\@citet@tag@n > 1}{
  \def\@citet@Tag{Tags}%
 }{%
  \def\@citet@Tag{Tag}%
 }%
 \@citet@Tag~\@citet@tags
}
\newcommand{\citet@sep}{, }
\newcommand{\citet@seplast}{ and }
\newcommand{\citet@entry}[2]{\citet@tag{#1}{#2}}
\let\@old@cite\cite
\renewcommand{\cite}[2][]{%
 \def\@citet@detail{\citet@taglist{#1}{\@citet@url}}%
 \ifthenelse{\equal{#2}{sp}}{%
  \def\@citet@url{\citet@url@sp}%
  \def\@citet@bib{\citet@bib@sp}%
 }{\ifthenelse{\equal{#2}{kd}}{%
  \def\@citet@url{\citet@url@kd}%
  \def\@citet@bib{\citet@bib@kd}%
 }{
  \def\@citet@detail{#1}%
  \def\@citet@bib{#2}%
 }}%
 \ifthenelse{\equal{#1}{}}{%
  \@old@cite{\@citet@bib}%
 }{%
  \@old@cite[\@citet@detail]{\@citet@bib}%
 }%
}
\makeatother

\newcommand{\desc}{\tu{desc}}

\newcommand{\hdesc}[2]{{#1\tu{-}\desc{\ifthenelse{\equal{#2}{}}{}{_#2}}}}

\DeclareMathOperator{\Li}{Li}
\newcommand{\cf}{\mathcal F}
\newcommand{\gp}[1]{{\ttfamily #1}}
\newcommand{\tabu}[1]{\begin{tabular}#1\end{tabular}}
\renewcommand{\tabl}[1]{\begin{table}#1\end{table}}

\begin{document}
\title[Nonexistence of GBF and quadratic norm form equations]{Nonexistence
 of generalized bent functions and the quadratic norm form equations}
\author[C. Lv]{Chang Lv}
\address{Key Laboratory of Cyberspace Security Defense\\
Institute of Information Engineering\\
Chinese Academy of Sciences\\
Beijing 100093, P.R. China}
\email{lvchang@amss.ac.cn}
\author[Y. Zhu]{Yuqing Zhu}
\address{Beijing Key Laboratory of Security and Privacy in Intelligent Transportation
	\\
	School of Cyberspace Science and Technology\\
Beijing Jiaotong University\\
Beijing 100044, P.R. China}
\email{zhuyuqing@bjtu.edu.cn}

\begin{abstract}
We present a new result on the nonexistence of  generalized bent functions (GBFs) from
 $(\ZZ/t\ZZ)^n$ to $\ZZ/t\ZZ$ (called type $[n,t]$) for a large class.
Assume $p$ is an odd prime number.
By showing certain quadratic norm form equations having no
 integral points,
 we obtain a universal result on the nonexistence of GBFs with type $[n, 2p^e]$
 when $p$ and $n$ satisfy  a certain inequality,
 and by computational methods with a widely accepted hypothesis, Generalized Riemann Hypothesis,
 we also achieve some results on the nonexistence of GBFs
 for relatively small $p$.
\end{abstract}
\maketitle


\section{Introduction}\label{sec_intro}
A Boolean function $\cf: (\ZZ/2\ZZ)^n \rightarrow \ZZ/2\ZZ$ is bent if and only if the discrete Fourier transform of its sign function $(-1)^\cf$ takes only two values $\pm 2^{n/2}$.
Bent functions are initially introduced by Rothaus~\cite{Rothaus} and deeply studied by him and Dillon~\cite{dillon1972survey,dillon1974elementary}.
Subsequently, a significant amount of research has been dedicated to this topic, resulting in the extension of the definition of bent functions in various manners.
This has given rise to several classes of generalized bent functions that share many of the useful properties of the original definition~\cite{BudaghyanCHKM12,Carlet93,CarletD04,CarletDL04,DobbertinLCCFG06,Kumar,MesnagerOS15,mesnager2016bent,mesnager2017decomposing,Nyberg91,OSW,feng2017complete}.

Bent functions are interesting combinatorial objects and have applications in design theory, coding theory and cryptography.
Bent functions are maximally nonlinear Boolean functions, which can introduce nonlinearity in the construction of stream ciphers and block ciphers.
However, bent functions are not balanced.
Directly using them will lead to a statistical correlation between the plaintext and the ciphertext.
Researchers generalized the definition of  bent to semi-bent, balanced (semi-) bent and partially bent
to enhance other cryptographic properties by slightly decreasing the nonlinearity.
For a more detailed history of bent functions, one can refer to \cite{CarletM16}.

Let $n\ge1$, $t\geq 2$ be two integers, $\ZZ/t\ZZ$ be the residue ring modulo $t$
 and $\zeta_t=\exp(2\pi\sqrt{-1}/t)$ be a primitive $t$-th root of unity.
 Kumar et al. \cite{Kumar} generalized the binary bent functions to the $t$-ary ones as follows.
\defi{\label{def_gbf}
A function $\cf: (\ZZ/t\ZZ)^n\ra \ZZ/t\ZZ$
 is called a \emph{Generalized Bent Function (GBF)} with \emph{type} $[n,t]$ if
\eq{\label{eq_F_abs}
 F(\lambda)\overline{F(\lambda)}=t^n
}
 for every $\lambda\in (\ZZ/t\ZZ)^n$, where
\eqn{
 F (\lambda)=\sum_{x\in(\ZZ/t\ZZ)^n}\zeta_t^{ \cf(x)}\cdot \zeta_t^{-x\cdot \lambda}
}
 is the discrete Fourier transform of the function $\zeta_t^{\cf(x)}$,
 $x\cdot \lambda$ is the standard dot product,
 and $\overline{F(\lambda)}$ is the complex conjugate of $F(\lambda)$.
}

In this paper, we focus on the nonexistence of GBFs.
For $t=2$, Rothaus \cite{Rothaus} proved that GBFs with type $[n,2]$ exist.
For $t>2$, Kumar et al. \cite{Kumar} constructed GBFs except the case where
 $n$ is odd and $t\equiv2\pmod 4$, with which type there is no GBF found until now.
Hence, we only consider the case where $n$ is odd and $t=2N$ with
$2\nmid N\ge3$.

Let $\varphi$ denote the Euler's totient function.
For $(a, N) = 1$, let $\ord_N(a)$ denote the order
of $a$ in the multiplicative group $(\ZZ/N\ZZ)^\tm$.
There are many results on the nonexistence of GBFs under some extra constraints:
\enmt{[(1)]
\item (Kumar \cite{Kumar})\label{it_kumar} type $[n, 2N]$ where $2\nmid N\ge3$,
 $2^s\equiv-1\pmod N$ for some  integer $s\ge1$, and $n$ is odd;
\item (Pei \cite{Pei}) type $[1,2\tm 7]$;
\item (Ikeda \cite{Ikeda})\label{it_ikeda}  type $[1, 2p_1^{e_1}\dots p_g^{e^g}]$
 where $p_1,\dots,p_g$ are distinct primes and \\
 $p_i^{s_i}\equiv-1\pmod{N/p_i^{e_i}}$ for some $s_i, i=1,\dots,g$;
\item (Feng \cite{feng-gbf1})\label{it_feng1} type $[n<m/s, 2p^l]$,
 where $n$ is odd, $p\equiv7 \pmod8$ is a prime,
 $s=\frac{\varphi(p^l)}{\ord_{p^l}(2)}$
 and $m$ is the smallest odd positive integer s.t. $x^2+py^2=2^{m+2}$ has
 integral solutions;
\item (Feng et al. \cite{feng-gbf1,feng-gbf2,feng-gbf3}) various classes with
 type $[n<m,2p_1^{l_1}p_2^{l_2}]$, where $n$ is odd, $p_1,p_2$ are two distinct
 primes satisfying certain conditions and $m$ is an upper bound for $n$;
\item (Jiang and Deng \cite{jiang-bent}) type $[3,2\tm 23^e]$;
\item (Li and Deng \cite{li-bent})\label{it_li}  type $[m,2p^e]$ where
 $p\equiv7 \pmod 8$ is a prime with $\ord_{p^e}(2)=\varphi(p^e)/2$
 and $m$ is the same as in \eqref{it_feng1};
\item (Lv and Li \cite{lv2017negbf})\label{it_lv_pq} type
 $[m,2p_1^{r_1}p_2^{r_2}]$ where $p_1\equiv7 \pmod 8$  and $p_2\equiv 5\pmod 8$
 are two primes satisfying some conditions  and $m$ is the smallest odd positive integer s.t. $x^2+p_1y^2=2^{m+2}$ has
 integral solutions;
\item (Lv and Li \cite{lv2017negbf})\label{it_lv_sp} type
 $[1\le n\le3,2\tm31^e]$ and
 $[1\le n\le5,2\tm151^e]$ where
 $e$ and $n$ are  positive integers and $n$ is  odd.
}

On the nonexistence of GBFs, some of the previous results were sporadic,
 some of them had too many constrains on parameters, and none of them included
 the cases of type $[n, 2N]$ with $N = p^e$, where $n, p, e$ are described in
 \eqref{eq_as_nontrivial}, i.e.,
 $p\equiv 1 (\bmod8)$ is a prime, $n\ge3$ and $f=\ord_p(2)$ is odd.

The  main result of this paper, Theorem \ref{thm_npe},
 provides a \emph{universal} result on the nonexistence for type $[n, 2p^e]$
 when $p$ and $n$ satisfy a certain inequality,
In particular, we partially solve the problem on the nonexistence of GBFs for
 the cases of \eqref{eq_as_nontrivial}.
In addition, we show that for a fixed $n$, there are infinitely many such $p$'s
 (Corollary \ref{cor_8_p}) under the Extended Riemann Hypothesis.
We also use a computational method with a similar hypothesis, Generalized Riemann Hypothesis,
 to give some results on the nonexistence  of GBFs for relatively small $p$.

The paper is organized as follows.
Basic results and facts needed in this paper are briefly introduced
 in Section \ref{sec_pre}.
Then in Section \ref{sec_norm} we study  the integral solvability of a class of
 quadratic norm form equations over subfields of cyclotomic fields,
 which is the main tool we shall use.
As an application, we prove Theorem \ref{thm_npe} in Section \ref{sec_npe}.
We also show that there are infinitly many primes $p$ that satisfy the conditions of the theorem.
Section \ref{sec_comp} is dedicated to some additional results on the nonexistence
 for relatively smalls $p$
 obtained by computational methods.

\section{Preliminaries} \label{sec_pre}

\subsection{Basic results on number theory}
Our methods for proving the nonexistence of GBFs with certain type involve algebraic
 number theory, such as cyclotomic fields and their subfields,
 ideals, class groups and Galois actions.
The standard references are \cite{janusz}  and \cite{Washington}.
Below we briefly introduce some basic results.

Assume $F$ is a number field. Denote by $\fo_F$ the ring of integers in $F$.
It is a Dedekind domain.
One can consider the \emph{fractional ideals} in $F$.
A fractional ideal is an $\fo_F$-module $\fa$ contained in $F$ such that there exists an element $\alpha \in \fo_F$ for which $\alpha \fa \subseteq \fo_F$.
Denote by $I_F$  the set of nonzero fractional ideals of $F$,
 which is a free abelian group  generated by the prime ideals  under multiplication.
A principal fractional ideal means a fractional ideal of the form
 $\beta\fo_F$ where $\beta\in F$.
The set of all nonzero principal fractional ideals, denoted by $P_F$,
 is a subgroup of $I_F$, and the quotient $I_F/P_F$, denoted by $Cl(F)$,
 is called the  \emph{class group} of $F$.
Class groups play an important role in classical algebraic number theory.
One of the nontrivial facts is that $Cl(F)$ is a finite abelian group for all $F$.
Denote by $h(F)$ the cardinality of $Cl(F)$, which is called the \emph{class number} of $F$.

We also need some basis results on the decompositions of prime ideals in  extension fields
 and the decomposition fields.
We only consider the unramified prime ideals in a cyclotomic extension.
For general cases, we refer the readers to
\cite[Section I.6, Section III.7]{janusz}.
Let $m$ be an  positive integer
with $m \not\equiv 2 \pmod 4$ and
 $K = \QQ(\zeta_m)$ be a cyclotomic field.
Then $K$ has degree $r = \varphi(m)$ over $\QQ$, where $\varphi$ denotes the Euler's totient function.
One has $\fo_K = \ZZ[\zeta_m]$.
For a prime $q\nmid m$, the ideal $q\fo_K$ will split into $g$ distinct
prime ideals, namely
\eqn{
q\fo_K = \prod_{i=1}^g \Q_i.
}
Since $K/\QQ$ is a Galois extension, the extension degree $[\fo_K/\Q_i: \ZZ/q\ZZ] = \ord_m(q)$
 is the same for each $i$,
 usually denoted by $f$, called \emph{relative degree} of $q$.
One has $r = fg$.
Since $\Gal(K/\QQ)$ is cyclic, there is a unique subfield $D\subseteq K$ having degree $g$ over $\QQ$
 called the \emph{decomposition field} of $q$, such that
\eqn{
 q\fo_D = \prod_{i=1}^g \q_i,
}
where $\q_i\fo_K = \Q_i$ and
$\q_i$ is a prime ideal in $D$ having relative degree $1$ over $\QQ$.

Now let us consider
 $K = \QQ(\zeta_{p^e})$, where $p$ is an odd prime and $e$ is a positive integer.
Since $\Gal(K/\QQ) \cong (\ZZ/p^e\ZZ)^\tm$ is cyclic,
 for any positive integer $s\mid [K:\QQ]$, there is a unique subfield of $K$ having degree $s$
 over $\QQ$.
Based on this fact, one can directly obtain the following two results:
One is that if $q$ is a prime distinct from $p$ and $f = \ord_{p^e}(q)$,
 the subfield  having degree $g = [K:\QQ]/f$ over $\QQ$ must be the
 decomposition field of $q$.
The other is that the unique quadratic subfield contained in $K$ is
  $\QQ(\sqrt{(-1)^{(p-1)/2}p})$.
Since the discriminant of $K$ is $\pm p^{p^{e-1}(pe-e-1)}$,
 we know that $p$ is the only ramified prime of $K$, and thus
 is the only ramified prime of the unique quadratic subfield contained in $K$.
By a direct calculation,
one can find  $\QQ(\sqrt{(-1)^{(p-1)/2}p})$ is the only candidate.

To determine the multiplicative order of $a$ modulo $p^e$, one can apply the following lemma.
\lemm{\label{lem_ordpe}
	Let $p$ be an odd prime and $a$ be an integer with $f=\ord_p(a)$.
	If \eq{\label{eq_afnmp2}
		a^f\not\equiv1\pmod{p^2},
	}
	then $\ord_{p^e}(a)=fp^{e-1}$ for all $e>1$.
}
\pf{
	One can easily obtain the lemma by following the strategy in
	\cite[Chapter 4.5]{hasse1978number} of proving that
	$1+p \bmod p^e$ generates a multiplicative subgroup of order $p^{e-1}$.
	For convenience of the readers,
	we provide a proof here.

	Since $f\mid p-1$ and $(p-1, p^{e-1})=1$, it suffices to show that
	\eq{\label{eq_equiv1}
		a^{fp^{e-1}}\equiv1\pmod{p^e}
	}
	and \eq{\label{eq_nequiv1}
		a^{fp^{e-2}}\not\equiv1\pmod{p^e}.
	}
	We claim that for each $r\ge2$, \eqn{
		a^{fp^{r-2}}\equiv p^{r-1}t+1\pmod{p^r}\text{ for some }
		t\in\ZZ\text{ not divisible by }p,
	}
	which implies \eqref{eq_nequiv1} when $r=e$ and
	\eqref{eq_equiv1} when $r=e+1$.

	We shall prove  this claim by induction on $r$.
	Since we have $a^f\equiv1\pmod p$ and \eqref{eq_afnmp2},
	the claim holds for $r=2$.
	Suppose it  holds for $r\ge2$. Then we have for some $k\in\ZZ$ that
	\aln{
		a^{fp^{r-1}}&=(p^rk+p^{r-1}t+1)^p\\
		&=(p^{r-1}t+1)^p+p(p^{r-1}t+1)^{p-1}p^rk+
		\binom{p}{2}(p^{r-1}t+1)^{p-2}(p^rk)^2+\dots+(p^rk)^p\\
		&\equiv(p^{r-1}t+1)^p=(p^{r-1}t)^p+\dots+\binom{p}{2}(p^{r-1}t)^2+p\tm p^{r-1}t+1\\
		&\equiv p^rt+1\pmod{p^{r+1}},
	}
	i.e., it also holds for $r+1$.
	This completes the proof for the lemma.
}

To determine the integral elements in quadratic extension fields, we have the following lemma.
\lemm{\label{lem_qr_int}
	Let $E=F(\sqrt{d})$ be an arbitrary quadratic extension of number fields,
	where $d\in \fo_F$ and $d\fo_F$ factors into a product of
	distinct prime ideals of $F$.
	Then every element of $\fo_E$ is of the form
	$(x+y\sqrt{d})/2$ for some $x,y\in\fo_F$.
}
\pf{
	For any $\beta\in \fo_E$, write $\beta=a+b\sqrt{d}$ where $a,b\in F$.
	We may assume $b\neq 0$. The minimal polynomial of $\beta$ over
	$F$ is \eqn{
		T^2-2aT+a^2-db^2\in F[T].
	}
	Since $\beta\in\fo_E$, we have $2a, a^2-db^2 \in \fo_F$.
	Hence $a=x/2$  and $db^2=z/4$ for some $x,z\in\fo_F$.
	Suppose the prime decomposition of $d$ is
	\eqn{
		d\fo_F=\p_1\dots\p_s,
	}
	and
	\eqn{
		2b\fo_F=\fa\p_1^{r_1}\dots\p_s^{r_s}
	}
	for some fractional ideal $\fa$ of $F$ coprime with
	$\p_1\dots\p_s$ and some integers $r_1,\dots,r_s$.
	Then we have
	\eqn{
		z = (2b)^2d\fo_F=\fa^2\p_1^{2r_1+1}\dots\p_s^{2r_s+1}\subseteq \fo_F.
	}
	It follows that $\fa\subseteq\fo_F$ and $r_1,\dots,r_s\ge0$,
	which implies $2b=y$ for some $y\in\fo_F$.
	The proof is complete.
}

We also need a result on
 asymptotic estimation of the number  of specific primes.
Let $N_a^{(n)}(x) = \{ p \le x \mid \ord_p(a) = (p-1)/n\}$.
\thmu{[{\cite[Thm. 1]{murata1991problem}}] \label{thmu_artin}
Let $a\geq 2$ be a squarefree integer and $n$ be a positive integer.
Assuming the
 Riemann Hypothesis, we have that for any $\epsilon> O$,
\eqn{
\sharp  N_a^{(n)}(x)  =
 C_a^{(n)} \Li (x) + O\left((n^\epsilon (\log\log x) + \log a)\frac{x}{\log^2x}
  \right),
}
where $\Li(x)=\int_2^x\frac{dt}{\ln t}$ and $C_a^{(n)}$ is a positive constant.
}

\subsection{GBFs and norm form equations}
Let $E/F$ be a quadratic extension of number fields.
An \emph{quadratic norm form equation} (norm equation for short) is a equation of the form
\eq{\label{eq_norm}
	N_{E/F}(\alpha)=a,
}
where $N_{E/F}$ is the norm from $E$ to $F$, $\alpha\in E$ is the  indeterminate and
$a\in F^\tm$ is a constant.

Assume $a\in\fo_F$.
We say $\alpha\in \fo_E$ is an \emph{integral point} of \eqref{eq_norm} if it
is a solution to \eqref{eq_norm},
where $\fo_E$ is the ring of integers of $E$.
In most cases, we obtain the explicit quadratic equation over $\fo_F$
\eqn{
	a_1x^2+a_2xy+a_3y^2+c=0,
}
where $a_i$ and $c$ are in $\fo_F$ for $i = 1, \dots, 3$ and the indeterminate
$(x,y)\in\fo_F^2$.
Equations of this form are
investigated in many papers, such as
\cite{wei_diophantine,lv2018intrepqr} for $F=\QQ$,
\cite{wei2015certain,wei1,wei2} for $F$ being quadratic fields, and
\cite{multi-norm-tori,multip-type} for $F$ being arbitrary number fields.

It was Feng \cite{feng-gbf1} that first used  the nonsolvability of the
 norm form equation
\eq{\label{eq_alpha2n}
\alpha\bar\alpha=(2N)^n,\quad\alpha\in\ZZ[\zeta_N]
}
 as the key idea to
 prove the nonexistence of GBFs with type $[n, 2N]$.
Feng  observed that
  if $E = \QQ(\zeta_N)$ and $t= 2N$,
  then
  \eqn{
  	F (\lambda)=\sum_{x\in(\ZZ/t\ZZ)^n}\zeta_t^{ \cf(x)}\cdot \zeta_t^{-x\cdot \lambda}
  }
  belongs to $\fo_E = \ZZ(\zeta_N)$, since $F (\lambda)$ is a linear combination of  $\zeta_t^i$ and $\zeta_t^i$ belongs to $\fo_E$.
  By definition,
\eqn{
F(\lambda)\overline{F(\lambda)}=(2N)^n.
}
It follows that $F(\lambda)$ is a solution to
the norm form equation~\eqref{eq_alpha2n}.

From then on, in order to prove the nonexistence of GBFs,
 various methods were developed
 (\cite{feng-gbf2, feng-gbf3, lv2017negbf}, etc.)
 to show the nonexistence of integral points of \eqref{eq_alpha2n}.
However, it is difficult to deal with the case where $N=p^e$ with $p\equiv1\pmod8$,
 and people know little about the nonexistence of GBFs with types in this case.
Our method in this paper, is rather different from any one before, and
 can deal with the less known cases.

We also use a technique called \emph{descent}, as described below.
\defi{[Self-conjugated \cite{Washington}]
Let $p$ be a prime integer, $m = p^lm'$ where $l \ge 0$ and $(p,m') = 1$.
We call $p$ to be \emph{self-conjugated}
 with respect to $m$ if there exists $s \in\ZZ$ such that $p^s\equiv - 1
  \pmod{m'}$.
}

\lemu{[{\cite[Lem. 2.4]{liu2016new}}] \label{lemu_descent}
Let $n \ge 2,\ m \ge 3$,
 $K = \QQ(\zeta_m)$.
Suppose that
 there exists $\alpha\in\fo_K$ such that $\bar \alpha\alpha = n$.
\enmt{
\item \label{it_pt} If $p$ is a common prime factor of $m$ and $n$,
 where $n = p^tn'$, $(p,n') = 1$ and $p$ is self-conjugated
 with respect to $m$, then there exists $\beta\in\fo_K$
  such that $\bar\beta\beta= n'$.
\item  If $q$ is a prime factor of $n$,
 where $n = q^an'$, $(q,n') = 1$, $(q,m) = 1$ and $q$ is self-conjugated with
 respect to $m$, then $a$ is even and there exists $\beta\in\fo_K$
 such that $\bar\beta\beta = n'$.
\item \label{it_ql} Suppose $n = q^l$ is a power of a prime number $q$ and
 $(q(q - 1),m) = 1$. Let $D$ be the
 decomposition field of $q$ in $K$.
 Then there exists $\beta\in\fo_K$ such that $\beta^2\in\fo_D$ and
  $\bar\beta\beta = n$.
}}

\section{Nonsolvability of quadratic norm form equations}\label{sec_norm}

\subsection{New results on nonsolvability of norm form equations}
Let $E$ be
 a complex subfield of the $N$-th cyclotomic field $\QQ(\zeta_N)$ and $F=E\cap\RR$ be its maximal
 real subfield. First, consider the case $N=p$, where $p$ is a prime.
Since $E$ is complex, we know that $E/F$ is quadratic.
We consider the nonexistence of
 integral points of the norm  equation \eqref{eq_norm} in
 the case where $a$ is a rational prime power $q^n$.
For several certain classes of $N$ and $q^n$,
 this problem was discussed in \cite{feng-gbf1,feng-gbf2,feng-gbf3,liu2016new,lv2017negbf,lv2017nepps}.
We give the following theorem.
\thm{\label{thm_norm_cm}
Let $p$ and $q$ be two distinct primes, $n$ be an odd positive integer
 and $K=\QQ(\zeta_p)$.
Assume $E\subseteq K$ is complex and $F=E\cap\RR$.
Suppose $[F:\QQ]=k$. If
\eq{\label{eq_as_pqnk}
p>(4q^n)^k,
}
then the quadratic norm form equation $N_{E/F}(\alpha)=q^n$
 has no solution with $\alpha\in\fo_E$.
}
\pf{
By assumption \eqref{eq_as_pqnk}, $p$ is an odd prime.
Let $\gamma=\zeta_p-\zeta_p^{-1}$.
We claim that $N_{K/\QQ}(\gamma)=p$.
Rewrite $\gamma$ as $\zeta_p(1-\zeta_p^{-2})$.
Note that $N_{K/\QQ}(1-\zeta_p)$ equals to the evaluation of $\prod_{i=1}^{p-1}(x-\zeta_p^i)=x^{p-1}+x^{p-2}+\cdots+1$ at 1.
Thus, $N_{K/\QQ}(1-\zeta_p)=p$.
Since $p$ is an odd prime, $\zeta_p \mapsto \zeta_p^{-2}$ is an automorphism.
We have  $N_{K/\QQ}(1-\zeta_p^{-2})=N_{K/\QQ}(1-\zeta_p)=p$.
Combining with the fact that $N_{K/\QQ}(\zeta_p)=1$ for any odd prime $p$,
we have $N_{K/\QQ}(\gamma)=p$.

Let $\xi=N_{K/E}(\gamma)$ and $\delta=N_{E/F}(\xi)$.
Thus we have
\eqn{
N_{E/\QQ}(\xi)=N_{K/\QQ}(\gamma)=p.
}
We claim that $E=F(\xi)=F(\sqrt{-\delta})$.
Actually, if $\xi\in F$,
\eqn{
p=N_{E/\QQ}(\xi)=N_{F/\QQ}(\xi^2)=N_{F/\QQ}(\xi)^2.
}
Since  $\gamma\in\fo_K$, $\xi = N_{K/E}(\gamma)\in \fo_E$, combining with the assumption $\xi\in F$, we have $\xi\in F\cap\fo_E = \fo_F$. It follows that
$N_{F/\QQ}(\xi)\in \ZZ$,
 which yields a contradiction.
This shows that $E=F(\xi)$.
Since $E$ is complex,
 we know that $[K:E]$ is odd (if $[K:E]$ is even,
 then $\Gal(K/E)$ would contain the complex conjugation,
 which is impossible since $E$ is complex)
 and $E/F$ is quadratic with Galois group generated by
 the complex conjugation.
Also note that $\Gal(K/\QQ)$ is abelian containing the complex conjugation.
Hence
\eqn{
\bar\xi=\overline{N_{K/E}(\gamma)}=N_{K/E}(\bar\gamma)=N_{K/E}(-\gamma)
 =-N_{K/E}(\gamma)=-\xi.
}
It follows that
\eqn{
\xi^2=-\xi\bar\xi=-N_{E/F}(\xi)=-\delta,
}
which completes the proof for the claim.

Thus $E=F(\sqrt{-\delta})$ with
\eqn{
N_{F/\QQ}(\delta)=N_{E/\QQ}(\xi)=p,
}
which implies that $\delta\fo_F$ is a prime ideal lying over $p$.

Now assume that
\eq{\label{eq_alpha}
N_{E/F}(\alpha)=q^n\text{ for some }\alpha\in\fo_E.
}
Due to Lemma \ref{lem_qr_int}, we may write
 $\alpha=(x+y\sqrt{-\delta})/2$ for some $x,y\in\fo_F$.
It follows from \eqref{eq_alpha} that
\eq{\label{eq_x2deltay2}
q^n=N_{E/F}(\alpha)=\alpha\bar\alpha=\frac{x^2+\delta y^2}{4}.
}
Since $n$ is odd, we have $y\neq0$. Otherwise, $x^2=4q^n$
 and then $\sqrt q\in F$, which yields a contradiction
 since $\QQ(\sqrt{(-1)^{(p-1)/2}p})$ is the
 unique quadratic subfield of $K=\QQ(\zeta_p)$.

Next we shall show that $x^2$, $y^2$ and $\delta$ are all \emph{totally nonnegative},
 i.e., they are all nonnegative after applying each $\s\in\Gal(F/\QQ)$.
Recall that  every element in $\Gal(K/\QQ)$ commutes with the complex conjugation.
We have for every $\s\in\Gal(K/\QQ)$ that
\eqn{
\s(\delta)=\s(N_{E/F}(\xi))=\s(\xi\bar\xi)=\s(\xi)\overline{\s(\xi)}\ge0.
}
Since $\s(x),\s(y)\in F$ are fixed by the complex conjugation, we have
\eqn{
\s(x^2)=\s(x)^2\ge0\text{ and }\s(y^2)=\s(y)^2\ge0.
}
It follows from \eqref{eq_x2deltay2}  that
\aln{
(4q^n)^k&=N_{F/\QQ}(x^2+\delta y^2)\\
 &=\prod_{\s\in\Gal(F/\QQ)}\left( \s(x^2)+\s(\delta)\s(y^2) \right)\\
 &\ge  \prod_{\s\in\Gal(F/\QQ)}\s(x^2) + \prod_{\s\in\Gal(F/\QQ)}\s(\delta)
     \prod_{\s\in\Gal(F/\QQ)}\s(y^2)\\
 &=N_{F/\QQ}(x)^2 +N_{F/\QQ}(\delta)N_{F/\QQ}(y)^2\\
 &\ge p,
}
where the first inequality holds since
 $x^2$, $y^2$ and $\delta$ are all totally nonnegative,
 and the second inequality holds since $N_{F/\QQ}(x)\in\ZZ$,
 $N_{F/\QQ}(y)\in \ZZ\setminus\{0\}$ and $N_{F/\QQ}(\delta)=p$.
This is a contradiction with the assumption \eqref{eq_as_pqnk}.
The proof is complete.
}

Now consider a special case of Theorem \ref{thm_norm_cm}
 where $E=K=\QQ(\zeta_p)$. In this case, $k=[F:\QQ]=(p-1)/2$ and
 the assumption \eqref{eq_as_pqnk} will not hold.
Fortunately, we may use \eqref{it_ql} in Lemma \ref{lemu_descent}
 to \emph{descent} Equation \eqref{eq_alpha}
from $\QQ(\zeta_p)$ to a subfield with a small degree over $\QQ$.
Then we obtain that there exists no integral point for \eqref{eq_norm}
 with a large class of $p$, $q$ and $n$.
We state the result after generalizing $\QQ(\zeta_p)$ to $\QQ(\zeta_{p^e})$.
For an integer $a$, denote by $\B(a)$ the $2$-part of $a$,
i.e., if $a=2^ma_1$ for some odd $a_1$, we have $\B(a)=2^m$.

\thm{\label{thm_norm_cyc}
Let $p$ and $q$ be two distinct primes
 with $f=\ord_p(q)>1$ and $n$ be an odd positive integer.
Suppose $e$ is an positive integer.
When $e>1$, we further assume that
\eq{\label{eq_qfnmp2}
q^f\not\equiv1\pmod{p^2}.
}
Let $E=\QQ(\zeta_{p^e})$ and $F=E\cap\RR$. If
\eq{\label{eq_as_pqnf}
p>4^{\B(l)}q^{nl}\text{ where }
  l=\frac{2(p-1)}{\left(3-(-1)^f\right)f},
}
then $N_{E/F}(\alpha)=\alpha\bar\alpha=q^n$
 has no solution with $\alpha\in\fo_E$.
}
\pf{
First note that $f>1$. Then $p$ is odd,  $E$ is complex and $E/F$
 is a quadratic extension.
Since $\Gal(E/\QQ)$ is cyclic of degree $\varphi(p^e)$,
 we denote by $E_q$  the unique  subfield of $E$ having degree
 $\frac{p-1}{f}$ over $\QQ$.
Then $E_q\subseteq K=\QQ(\zeta_p)$.
Assume there exist $\alpha\in \fo_E$ such that
\eqn{
N_{E/F}(\alpha)=\alpha\bar\alpha=q^n.
}
Note that $f=\ord_p(q)>1$ implies $p\nmid q-1$.
Since we have $q^f \not \equiv 1 \pmod{p^2}$ when $e>1$, we have $\ord_{p^e}(q) =fp^{e-1}$ for any positive integer $e$ according to Lemma \ref{lem_ordpe}.
It follows that
\eqn{
\frac{\varphi(p^e)}{\ord_{p^e}(q)}=\frac{\varphi(p)}{\ord_p(q)}=\frac{p-1}{f}.
}
Hence $E_q$ is the decomposition field of $q$ in $E$.
According to \eqref{it_ql} in Lemma \ref{lemu_descent}, we obtain that
\eq{\label{eq_beta}
\beta\bar\beta=q^n\text{ for some }\beta\in\fo_E\text{ and }\beta^2\in\fo_{E_q}.
}
It means that $\beta\in \fo_{E_1}$ where $E_1/E_q$ is some extension
 contained in $E$ such that  $[E_1:E_q]\le2$.

If $E_1$ is real, we have $\bar\beta=\beta$ and then
 $\sqrt q\in E$, which is impossible.
Thus $E_1$ is complex. Note that  $[E:E_q]=\ord_{p^e}(q)=fp^{e-1}$.
If $f$ is odd, we have $E_1=E_q$ and then $[E_1:\QQ]=(p-1)/f$.
Otherwise, since $[K: E_q]=f$ is even and $E/\QQ$ is cyclic,
 we may fix $E_1\subseteq K$ to be the unique quadratic extension of $E_q$
 and then we have $[E_1:\QQ]=2(p-1)/f$.
It follows that we always have \eqn{
[E_1:\QQ]=2\frac{2(p-1)}{\left(3-(-1)^f\right)f}=2l \text{ and $E_1\subseteq K$.}
}
We want to obtain a complex subfield $E_2\subseteq E_1$
 as small as possible.
Let $E_2\subseteq E_1$ be the unique subfield having degree
 $2\B(l)$ over $\QQ$.
Since $E_1$ is complex and  $[E_1:E_2]=l/\B(l)$ is odd,
 we have that $E_2$ is complex. 
 Let $F_2=E_2\cap\RR$, which has degree
 $\B(l)$ over $\QQ$.
 Taking norm from $E_1$ to $E_2$  in \eqref{eq_beta}, we obtain
\eq{\label{eq_nbeta}
N_{E_2/F_2}(N_{E_1/E_2}(\beta))=
 N_{E_1/E_2}(\beta)\overline{N_{E_1/E_2}(\beta)}
 =q^{nl/\B(l)}\text{ with }
 N_{E_1/E_2}(\beta)\in\fo_{E_2}.
}
Note that $E_2\subseteq K=\QQ(\zeta_p)$.
Substituting $E/F,\alpha,k$ and $n$ in  Theorem \ref{thm_norm_cm} by $E_2/F_2, N_{E_1/E_2}(\beta), \B(l)$ and $nl/\B(l)$ respectively, we have that
 Equation \eqref{eq_nbeta} does not hold
 due to the assumption \eqref{eq_as_pqnf}.
This completes the proof.
}

\subsection{Asymptotic estimations on the density of $p$'s}
Let us make the assumptions in Theorem~\ref{thm_norm_cyc} more explicit.
We will show that for fixed $q$ and $n$,
 there are infinitely many $p$'s satisfy all the  assumptions.

First we consider  Theorem~\ref{thm_norm_cyc}  in the case
 where $e=1$, i.e., $E=\QQ(\zeta_p)$.
\prop{\label{prop_den_artin}
Let $q$ be a prime number and
 $n$, $g$ are two positive integers.
For a positive real number $x$, let  $\pi(x)$ be the number of primes not exceeding $x$
 and $M_{q^n,g}(x)$ be the number of
 primes $p$ not exceeding $x$, such that
\enmt{[{\upshape(a)}]
\item $(p-1)/f=g$, where $f=\ord_p(q)>1$,\label{den.it_artin}
\item $p>4^{\B(l)}q^{nl}$, where
 $l=\frac{2(p-1)}{\left(3-(-1)^f\right)f}$. \label{den.it_p}
}
In particular, $p$ meets
 assumption \eqref{eq_as_pqnf} in Theorem~{\upshape \ref{thm_norm_cyc}}
 for the case $e = 1$.
Assuming \emph{Extended Riemann Hypothesis (ERH)}, then
\eqn{
M_{q^n,g}(x) \sim C_{q^n,g} \pi(x)
 \sim C_{q^n,g} \frac{x}{\log x}, \quad\text{ as }x\ra +\infty,
}
 where  $C_{q^n,g}$ is a positive constant that only depends on $q^n$ and $g$.
}
\pf{
The first constraint \eqref{den.it_artin} on $p$ is related to
 a kind of  generalization of  \emph{Artin's conjecture on primitive roots}.
Note that $f>1$ is automatic for sufficiently large $p$.
By Theorem \ref{thmu_artin}, where we take $n=g$ and $a=q$,
 the number of primes $p$ not exceeding $x$ and satisfying \eqref{den.it_artin}
 equals asymptotically to $C_q^{(g)}\Li(x)$ as $x\ra+\infty$ under ERH,
 where $C_q^{(g)}$ is a positive constant that only depends on $q$ and $g$.
As for the constraint \eqref{den.it_p},
 it suffices to exclude finitely many
 $p\le(4q^n)^g$ since $(4q^n)^g\ge4^{\B(l)}q^{nl}$.
It follows that for some positive $C_{q^n,g}$ only depending on $q^n$ and $g$,
\eqn{
M_{q^n,g}(x) \sim C_q^{(g)}\Li(x) \sim C_{q^n,g} \pi(x),\quad\text{ as }x\ra +\infty,
}
since  by the \emph{prime number theorem} we have
\eqn{
\Li(x)\sim \pi(x)\sim \frac{x}{\log x}.
}
This completes the proof.
}

Next we consider the case where $e>1$ and $E=\QQ(\zeta_{p^e})$ in Theorem~\ref{thm_norm_cyc}.
Besides  \eqref{den.it_artin} and \eqref{den.it_p}, there  is one more constraint
 $q^f\not\equiv1\pmod{p^2}$. We tighten it to
\enmt{[(a)]
\setcounter{enumi}{2}
\item $q^{p-1}\not\equiv1\pmod{p^2}$. \label{den.it_wieferich}
}

Note that primes NOT satisfying \eqref{den.it_wieferich} are called \emph{base-$q$ Wieferich primes}.
It is conjectured that the number of
 base-$q$ Wieferich primes not exceeding $x$ equals asymptotically to
 $C_q\log\log x$ as $x\ra+\infty$ \cite{murata1981average}, where $C_q$ is a constant only depending on $q$.
Since $\log\log x$ is negligible compared to $\pi(x)$,
 based on this conjecture and Proposition \ref{prop_den_artin}, we
 propose the following conjecture.
\con{\label{conj_den_wieferich}
With $q^n$ and $g$ fixed as in Proposition {\upshape \ref{prop_den_artin}},
 the number of  primes not exceeding $x$, such that
 the conditions \eqref{den.it_artin}, \eqref{den.it_p}
 and \eqref{den.it_wieferich} hold $($in particular, $p$ meets
 assumptions \eqref{eq_qfnmp2} and
 \eqref{eq_as_pqnf} in Theorem~{\upshape \ref{thm_norm_cyc}}
 for $e > 1)$
 equals asymptotically to
\eqn{
C_{q^n,g} \frac{x}{\log x}\text{ as }x\ra +\infty,
}
 where  $C_{q^n,g}$ is a positive constant that only depends on $q^n$ and $g$.
}

In fact, computational evidence indicates that Wieferich primes are extremely rare,
and one can see Remark \ref{rk_8_pe} \eqref{8_pe.it_wieferich} in the next section.

\section{Results on the nonexistence for GBFs with type $[n, 2p^e]$} \label{sec_npe}
\subsection{Main results}
By applying Theorem~\ref{thm_norm_cyc} we obtain the following theorem.
\thm{ \label{thm_npe}
Assume $p$ is an odd prime number.
Let $f=\ord_p(2)$ and $N=p^e$, where $e$ is a positive integer.  When $e>1$, we further assume that
\eq{\label{eq_2fnmp2}
2^f\not\equiv1\pmod{p^2}.
}
Let $n$ be an odd positive integer. If
\eq{\label{eq_as_p2nf}
p>2^{2\B(l)+nl}\text{ where }
  l=\frac{2(p-1)}{\left(3-(-1)^f\right)f},
}
then there is no GBF with type $[n,2N]$.
}
\pf{
Assume that $\cf$ is a GBF with type  $[n, 2N]$.
Since $f=\ord_p(2)$, we have $f>1$.
Let $E=\QQ(\zeta_N)=\QQ(\zeta_{p^e})$ and $F=E\cap \RR$.
By \eqref{eq_F_abs} in the definition of GBFs, we have
 that $F(\lambda)\in \fo_E$  and
\eqn{
F(\lambda)\overline{F(\lambda)}=(2N)^n = 2^np^{en}.
}
By \eqref{it_pt} in
 Lemma \ref{lemu_descent}, 
 we have
\eq{\label{eq_alpha2}
\alpha\bar\alpha=2^n
}
 for some $\alpha\in \fo_E$.

Take $q=2$ in Theorem~\ref{thm_norm_cyc}.
Then the assumptions in the theorem are fulfilled.
 We obtain that \eqref{eq_alpha2} has no solution with $\alpha\in\fo_E$, which yields a contradiction.
 This completes the proof.
}

Roughly speaking, Theorem \ref{thm_npe} provides that there is no GBF with type $[n,2p^e]$
 when $p$ is extremely larger than $n$ and $(p-1)/\ord_p(2)$.
By the discussion at the end of Section \ref{sec_norm} we are able to show that there are infinitely many primes
 $p$ satisfying the assumptions
 in Theorem \ref{thm_npe},
 for a fixed $n$.
\cor{\label{cor_8_p}
Let $n$ and $g$ be  fixed  positive integers.
Assuming ERH, then
 there exist a positive constant $C_{n,g}$  only depending on $n$ and $g$,
 such that as $x$ goes to infinity, there are asymptotically at least
 $C_{n,g} x/\log x$  primes $p$ not exceeding $x$ satisfying
 \[
 (p-1)/f=g  \text{ and } p>2^{2\B(l)+nl}
  ,
 \]
where $f=\ord_p(2)$ and $l=\frac{2(p-1)}{\left(3-(-1)^f\right)f}$, and for such $p$'s there is no GBF with type $[n,2p]$.

In particular, for each $n$, there are infinitely many such $p$'s.
}
\pf{
Applying Proposition \ref{prop_den_artin} with $q=2$,
 we obtain that asymptotically there are at least
 $C_{n,g} x/\log x$  primes $p$ not exceeding $x$, which satisfy all the assumptions
 of Theorem \ref{thm_npe} with $e=1$. The result then follows.
}

\rk{\label{rk_8_pe}
\enmt{[(a)]
\item  The assumptions in Theorem \ref{thm_npe} can be verified by direct calculations.
\item \label{8_pe.it_erh} Note that ERH is always considered true
 in computational practice.
\item \label{8_pe.it_wieferich} Moreover, if we assume
 the conjecture on asymptotic number of Wieferich primes (based-$2$)
 holds, then Conjecture \ref{conj_den_wieferich} tells us that Corollary \ref{cor_8_p} is
 also correct for $e>1$. That is, there are asymptotically at least
 $C_{n,g} x/\log x$  primes $p$ not exceeding $x$ such that
 there is no GBF with type $[n,2p^e]$.
In fact, numerical evidence  suggests that
 few primes in a given interval are Wieferich primes.
The base-$2$ Wieferich primes currently known are only $1093$ and $3511$,
 and they are the only two for all primes less than $6.7\tm 10^{15}$
  \cite{dorais2011wieferich}.
}}

For GBFs with type $[n,2p^e]$, where $n$ is odd and $f=\ord_p(2)$,
 the known results \eqref{it_kumar}, \eqref{it_ikeda} and \eqref{it_feng1}
 in the introduction cover the case $p\not\equiv1\pmod8$, the case
 $n=1$ and the case $f$ is even.
Therefore,
it is more significant  to
 restrict  Theorem \ref{thm_npe} to the case where $p\equiv1\pmod8$,
 $n\ge3$ and $f$ is odd.
 In this case, none of the method appearing in the previous literature are applicable.

Subsequently, we present several examples to illustrate the exact implications of Theorem \ref{thm_npe}
 in the case where
\eq{\label{eq_as_nontrivial}
\text{$p\equiv1(\bmod8)$, $n\ge3$ and $f=\ord_p(2)$ is odd.}
}

\

\subsection{Examples}
We give some examples based on the previous discussion.
The numerical results in these examples are all new and the calculations involved are all  elementary.

We first consider a fixed $p$ under the assumption \eqref{eq_as_nontrivial}.
\egu{\label{eg_fix_p}
By \eqref{eq_as_p2nf}, $l$ reaches the smallest possible value $4$
 when  $(p-1)/f=8$. Since $n\ge3$,
\eq{\label{eq_first_p}
p>2^{2\B(l)+nl}\ge2^{2\B(l)+3l}=2^{20}=1048576.
}
Thus  we search for primes $p>2^{20}$ satisfying \eqref{eq_as_nontrivial},
 of which the smallest  is $1049177$ and we obtain the smallest nontrivial
 instance, i.e., there is no GBF with type $[3, 2p]$ for  $p=1049177$.

In the same manner, by slightly increasing $n$, say $11$ and $15$,
 we obtain the nonexistence
 of GBFs with type $[n, 2p]$ for \aln{
&\text{ odd $n\le11$ with $p=4503599627370889$ and}\\
&\text{ odd $n\le15$ with $p=295147905179352827401$.}
}}

Next, we consider a fixed $n$ under the assumption \eqref{eq_as_nontrivial}.
\egu{\label{eg_fix_n}
Let $n=3$. By \eqref{eq_first_p}, we  search for $p$ satisfying \eqref{eq_as_p2nf} and
 obtain the first $5$ such primes:
 $1049177$, $1050169$, $1050233$, $1050473$, $1051961$.
For these primes $p$,  there are no GBFs with type $[3, 2p]$.
In fact, there is no GBF with type $[3, 2p]$ for any $p\ge1049177$
 such that $(p-1)/f=8$, and 
 they are infinitely many such $p$'s under ERH,
 by Corollary \ref{cor_8_p}.

We give another instance for $n=17$. By searching for primes $p$ satisfying \eqref{eq_as_p2nf}, we obtain
\eqn{
p=75557863725914323420409,\ 75557863725914323422233,\  \ldots
}
For each $p$ and any odd number $n\le17$
 there is no GBF with type $[n, 2p]$.
In the same manner,  we find that
 there is no GBF with type $[n, 2p]$ for all $p\ge2^{76}$ such that $(p-1)/f=8$.
}

At last we consider the case where $e>1$, under the
 assumption \eqref{eq_as_nontrivial}.
\egu{
 For $e>1$, to show the nonexistence of GBFs we need additional assumption \eqref{eq_2fnmp2}.
 By \cite{dorais2011wieferich}, except $1093$ and $3511$,
 all the primes $p$ less than  $6.7\tm 10^{15}$ satisfy
 assumption \eqref{eq_2fnmp2} (see Remark \ref{rk_8_pe} \eqref{8_pe.it_wieferich}).
Note that neither $1093$ nor $3511$ is congruent to $1$ modulo $8$.
It follows that for all positive integer $e$,
 there is no GBF with type $[n, 2p^e]$ for the $(n,p)$ pairs described in the previous two examples.
}

\section{Results on the nonexistence for relatively small $p$}\label{sec_comp}
Theorem \ref{thm_npe} in the previous section provides the nonexistence of GBFs with type $[n,2p^e]$
for a class of large primes $p$.
The smallest nontrivial instance  (see Example \ref{eg_fix_p}) is that
 there is no GBF with type $[3, 2p]$ for $p=1049177$.
In fact, according to \eqref{eq_as_p2nf}, the smallest prime $p$ satisfying the conditions in the theorem
grows exponentially with respect to $n$.

Thus in this section, we give a computational approach to deal with the case where $p$ is relatively small.
We first give a necessary condition that a GBF must meet, see Proposition~\ref{prop_sp_general}.
Then we give some results on the nonexistence  of GBFs with certain types by verifying the condition.

\subsection{Conditions on the exponents of prime ideals}
We proceed with more general parameters.
Assume $n$ is an odd number.
Let $t=2N$ with $2\nmid N\ge3$, $K=\QQ(\zeta_N)$,
 $f=\ord_N(2)$ and $g=\varphi(N)/f$.
Let $E$ be the decomposition field of $2$ in $K$. Then $[E:\QQ]=g$.
We always assume $f$ is odd here.

Now we suppose there is a GBF with type $[n, t=2N]$.
Then the same argument as in the beginning of
 the proof for Theorem \ref{thm_npe} yields
\eqn{
\alpha\bar\alpha=2^n
}
 for some $\alpha\in \fo_K$.
Next  by
 Lemma \ref{lemu_descent} \eqref{it_ql}
 we have
\eqn{
\beta\bar\beta=2^n\text{ for some }\beta\in\fo_K\text{ and }\beta^2\in\fo_E.
}
Since $[K:E]=f$ is odd, we know  that $g$ is even, $\beta\in\fo_E$ and
 $E$ is complex.
Thus we may assume the  prime decomposition of $2$ in $E$ is
\eqn{
	2\fo_E=\fP_1\fP_2\dots\fP_g=\fP_1\fP_2\dots\fP_u\bar\fP_1\bar\fP_2\dots\bar\fP_u,
}
where $\fP_k$'s are prime ideals in $E$ and $\fP_{u+k}=\bar\fP_k,\ k=1,2,\dots,u$ with $u=g/2$.
Then we have \eqn{
\beta\bar\beta\fo_E=\prod_{j=1}^u\fP_j^n\bar\fP_j^n.
}
Hence,
\eq{\label{eq_beta_dec}
\beta\fo_E=\prod_{j=1}^u\fP_j^{n_j}\bar\fP_j^{\bar n_j},
}
 where $n_j, \bar n_j$ are nonnegative integers such that
 $n_j+\bar n_j=n$ for all  $j=1,2,\dots,u$.
Since the left-hand side of the above equation is a principal ideal, we have the following equation
\eq{\label{eq_orig_rel}
	\prod_{j=1}^u\{\fP_j\}^{n_j}\{\bar\fP_j\}^{\bar n_j}=1,
}
where $\{\fP_j\}$ represents the class of $\fP_j$ in $Cl(E)$.

\propu{\label{prop_sp_general}
With the above notation, if  there are no  nonnegative integers
\[
(n_1,n_2,\dots,n_u,\bar n_1,\bar n_2,\dots,\bar n_u), \text{ where }n_j+\bar n_j=n\text{ for } j=1,2,\dots,u,
\]
such that Equation~\eqref{eq_orig_rel} holds,
 then there is no GBF with type $[n, 2N]$.
}

Subsequently, we mainly focus on the cases where $N=p^e$
 and the methods in previous literature are not applicable.
That is, we work under the assumption
\eq{\label{comp.eq_as_nontrivial}
\text{$p\equiv1(\bmod8)$, $n\ge3$ and $f=\ord_p(2)$ is odd.}
}
If $e>1$, we further assume that
\eq{\label{comp.eq_2fnmp2}
2^f\not\equiv1\pmod{p^2}
}
as in Theorem \ref{thm_npe}; then it reduces to the case where $e=1$.
\rk{\label{rk_wieferich}
Since neither $1093$ nor $3511$ is congruent to $1$ modulo $8$,
 \eqref{comp.eq_2fnmp2} holds for all $p$ less than  $6.7\tm 10^{15}$.
See Remark \ref{rk_8_pe} \eqref{8_pe.it_wieferich}.
}

\subsection{Algorithms and results}
Let $K = \QQ(\zeta_p)$ and $E$ be the unique subfield of $K$
 having degree $g$ over $\QQ$.
Proposition \ref{prop_sp_general} provides
 an explicitly algorithm to find $N=p$ and $n$, which enumerates
 $(n_1,n_2,\dots,n_g)$  to determine the solvability of
 \eqref{eq_orig_rel}.
This is implemented by GP \cite{pari} as follows.
\enmt{[\indent Step i.]
\item Given $p$ and $n$, use \gp{galoissubcyclo} to obtain the polynomial
 for the subfield  $E\subseteq K$, \gp{bnfinit} the field information of $E$
 involving the ideal class group $Cl(E)$,
 and \gp{idealprimedec} the set of primes $S=\{\fP_1, \dots, \fP_g\}$.
 \label{it_bnfinit}
\item Use \gp{nfgaloisconj} and \gp{nfgaloisapply} to identify the complex
 conjugation and the conjugate pairs of primes in $S$. Then we may assume
 $\fP_{u+k}=\bar\fP_k,\ k=1,2,\dots,u$.
\item Enumerate $n_1, \dots, n_u$ in the range $[0, n]$, calculate every ideal
 $\A=\prod_{j=1}^u\fP_j^{n_k}\bar\fP_j^{\bar n_j}$, and use \gp{bnfisprincipal}
 to see whether  $\A$ is principal, i.e.,
 to determine the solvability of \eqref{eq_orig_rel}.
 \label{it_search}
}

\rk{
Let us give some remarks on the implementation above.
\enmt{[(a)]
\item In Step \ref{it_bnfinit}, we can only deal with small $g$, otherwise
 the calculation of the field information will cost too much time and memory.
\item As noted in the PARI/GP documentation \cite{pari},
 since we use \gp{bnfinit} to calculate the class group, all results
 rely on this implementation
 are valid only under \emph{Generalised Riemann Hypothesis (GRH)}.
But as in  Remark \ref{rk_8_pe} \eqref{8_pe.it_erh},
 the results could be considered unconditionally
 correct in practice.
\item From Step \ref{it_search} we know that the size of the searching space,
 $O((n+1)^{\frac{g}{2}})$, predominantly influences the calculation time.
}}

According to the analysis above, we  only confine  ourselves to the case
 where $g=8$ and  $n$ is small.
 For each prime $p$ less than $3000$ with $\ord_p(2)=(p-1)/8$ being odd, we compute  the largest positive
 odd integer $n_p$ such that  \eqref{eq_orig_rel} is not solvable for all
 odd positive $n\le n_p$.
It costs several hours in an ordinary computer with Intel(R) Core(TM) 2 CPUs
 and 2G memory.
 Table~\ref{tab_pnp} lists these $(p,n_p)$.
By Proposition \ref{prop_sp_general} and Remark \ref{rk_wieferich}, we have
 the following proposition.
\propu{
For all $(p, n_p)$ in Table \ref{tab_pnp},
 there is no GBF with type $[n,2p^e]$ for any
 odd positive $n\le n_p$ and any positive integer $e$, under GRH.
}
\tabl{[htbp]
\caption{$(p, n_p)$ with $p<3000$ and $\ord_p(2)=(p-1)/8$ being odd}
\label{tab_pnp}
\centering
\tabu{{|r|r||r|r|}
 \hline
 $p$ & $n_p$ & $p$ & $n_p$ \\
 \hline
 89 & 3 & 1609 & 23 \\
 233 & 7 & 1721 & 19 \\
 937 & 7 & 1913 & 25 \\
 1289 & 13 & 2441 & 31 \\
 1433 & 17 & 2969 & 33 \\
 \hline
}}

Recall that Example \ref{eg_fix_n} shows that
 there is no GBF with type $[3, 2p]$ for all $p\ge1049177$
 and $(p-1)/f=8$.
Thus we use the previous implementation to determine the solvability
 of \eqref{eq_orig_rel} for  all $p<1049177$ and
 $n\le3$ such that  $f=(p-1)/8$ is odd.
The calculation also costs several hours and it shows that
 \eqref{eq_orig_rel} is not solvable for $n\le3$ and these $p$'s.
Combining with Proposition \ref{prop_sp_general},
 Example \ref{eg_fix_n} and Corollary \ref{cor_8_p},
 we obtain the following proposition.
\propu{ \label{propu_3}
Let $f=\ord_p(2)$. Under GRH,
 there is no GBF with type $[3, 2p^e]$ for all prime $p$ such that
 $f=(p-1)/8$ is odd and $\ord_{p^e}(2)=fp^{e-1}$.

}

\rk{
In Proposition \ref{propu_3},
 for the case $e>1$, the equality $\ord_{p^e}(2)=fp^{e-1}$ can be easily checked by testing
if $2^f \not \equiv 1 \pmod {p^2}$, see Lemma~\ref{lem_ordpe}.
}

\rk{
\enmt{
\item The first author and Li \cite{lv2017negbf} suggest that
 one should investigate the relations between
 the primes $\fP_k$ lying over $2$ in the class group $Cl(E)$, where
 $E$ is the decomposition field of $2$ in $\QQ(\zeta_p)$.
Unfortunately, here we cannot make a similar theoretical analysis as
 in  \cite{lv2017negbf}.
This is because under the assumption \eqref{eq_as_nontrivial}, $(p-1)/\ord_p(2)$, the degree of $E$
 over $\QQ$ (i.e., the number of $\fP_k$), is at least $8$.
Hence it becomes difficult to deal with the relations between $\fP_k$ in $Cl(E)$.
This is why we consider a computational approach.

\item
Note that the implementation in this section can deal with any integer $N$, not limited to the case where $N=p^e$.
This allows us to obtain more 
 computational results on the nonexistence of GBFs in the future.
}}

\section*{Acknowledgment}
We would like to thank  the referees for their valuable comments and suggestions.
We are also grateful to Yupeng Jiang, Jianing Li and Baofeng Wu for many helpful discussions.

\bibliography{unibib}
\bibliographystyle{amsalpha}
\end{document}